\documentclass[useAMS,usenatbib]{mn2e}

\usepackage{graphicx} 
\usepackage{multirow}
\usepackage{array}

\usepackage{amsmath,amssymb}
\usepackage{color}  

\newcommand{\bs}[1]{\boldsymbol{#1}}

\newcommand{\cs}{c_{\rm s}}
\newcommand{\kB}{k_{\textsc{b}}}
\newcommand{\rc}{r_{\rm c}}
\newcommand{\vc}{v_{\rm c}}
\newcommand{\vci}{v_{\rm c,\infty}}
\newcommand{\ud}{\textrm{d}}
\newcommand {\apgt} {\ {\raise-.5ex\hbox{$\buildrel>\over\sim$}}\ }
\newcommand {\aplt} {\ {\raise-.5ex\hbox{$\buildrel<\over\sim$}}\ } 
\DeclareMathOperator{\Li}{Li}

\usepackage[T1]{fontenc}
\usepackage[latin9]{inputenc}
\usepackage{float}
\usepackage{graphicx}
\usepackage{esint}

\makeatletter

\DeclareFontEncoding{LGR}{}{}
\DeclareTextSymbol{\~}{LGR}{126}

\begin{document}

\title{Ruling Out Bosonic Repulsive Dark Matter in Thermal Equilibrium}










\author[Slepian and Goodman]{Zachary Slepian\thanks{E-mail: zslepian@cfa.harvard.edu} and Jeremy Goodman\thanks{E-mail: jeremy@astro.princeton.edu}\\ 
Department of Astrophysical Sciences, Princeton University, Princeton, NJ, 08544\\
}
\maketitle

\begin{abstract}
Self-interacting dark matter (SIDM), especially bosonic, has been considered a promising
candidate to replace cold dark matter (CDM) as it resolves some of the problems associated
with CDM.  Here, we rule out the possibility that dark matter is a repulsive boson in
thermal equilibrium.  We develop the model first proposed by Goodman (2000) and derive the
equation of state at finite temperature.  Isothermal spherical halo models indicate a
Bose-Einstein condensed core surrounded by a non-degenerate envelope, with an abrupt
density drop marking the boundary between the two phases.  Comparing this feature with
observed rotation curves constrains the interaction strength of our model's DM particle, and
Bullet Cluster measurements constrain the scattering cross section.  Both ultimately can
be cast as constraints on the particle's mass.  We find these two constraints cannot be
satisfied simultaneously in any realistic halo model---and hence dark matter cannot be a
repulsive boson in thermal equilibrium.  It is still left open that DM may be a repulsive
boson provided it is not in thermal equilibrium;
this requires that the mass of the particle be significantly less than a millivolt.

\end{abstract}

\begin{keywords}
dark matter, galaxies: haloes, cosmology: theory, cosmology: observations

\end{keywords}

\section{Introduction}
There is much observational evidence for dark matter---$M/L$ ratios in low surface brightness galaxies (LSB's), flat rotation curves in spiral galaxies, the Oort discrepancy, gravitational lensing, cluster gas masses, the cosmic microwave background (CMB) combined with Big Bang nucleosynthesis (BBN) (Komatsu et al. 2011, Bertone et al. 2005, Sahni 2004, Peebles 1993)---but its essence remains elusive.  The standard model has been cold, non-interacting dark matter (CDM), which, while successful in simulations of large-scale structure formation, is less so in the details.  Steep density profiles (Colin et al. 2004) (though better baryon physics may improve this: see Cole et al. 2011 and Jardel \& Sellwood 2009), overproduction of small halos (Mikheeva et al. 2007, Diemand et al. 2005), failure to predict the zero point of the Tully-Fisher relation (Mo \& Mao 2000), and spiral galaxy bar rotation speeds (De Battista \& Sellwood 2000, 1998) all present unresolved questions for CDM.  

\newpage

In the early aughts, self-interacting dark matter (SIDM) appeared appealing because it promised to resolve some of these questions. A spate of papers explored SIDM (Spergel \& Steinhardt 2000, Goodman 2000, Wandelt 2000, Peebles 2000) and of late interest has  returned to this topic (Loeb \& Weiner 2011, Su \& Chen 2011).  In particular, bosonic SIDM has provoked considerable theoretical work (Chavanis 2011, Rindler-Daller \& Shapiro 2011, Rindler-Daller \& Shapiro 2010), especially as scalar-field dark matter (SFDM)  (the scalar field encodes a self-interaction) (Briscese 2011, Amaro-Seoane et al. 2010, Lee 2009, Lee \& Lim 2010, Urena-Lopez 2009, Bernal et al. 2008, Bernal \& Guzman 2006, Matos \& Urena-Lopez 2002).  We also note that many have studied non-self-interacting bosonic dark matter; see Harko 2011 I and II, Bernal et al. 2008, and Urena-Lopez 2009.  Non-bosonic self-interacting dark matter has also been considered; see Koda \& Shapiro 2011, Ahn \& Shapiro 2005, Mitra 2005, and Hennawi \& Ostriker 2001.

Given this, it is worthwhile to consider whether SIDM could produce realistic halo density profiles and rotation curves, with the aim of moving away from these models if they cannot.  In this paper, we calculate density profiles and rotation curves for a representative such model first presented in Goodman 2000.  The dark matter is a short-range repulsively interacting boson with repulsion strength, mass, and interaction cross section to be determined by observational constraints.  We place an upper bound on the DM particle mass using a constraint from the Bullet Cluster, and a lower bound on the mass by demanding that the dark matter halo, modeled as an isothermal sphere, produce an observationally allowed rotation curve. Note that this second constraint applies only if the DM is in thermal equilibrium (further discussion in \S5).  

These two constraints are incompatible: the lower bound is greater than the upper bound.
Hence we conclude that bosonic repulsive dark matter (RDM) in thermal equilibrium can be
ruled out.  This conclusion can be avoided if RDM halos have not reached even local
thermodynamic equilibrium by the present epoch.  However, in that case the model is very
similar to the traditional, effectively non-interacting axion (because the scattering must
be small enough so as not to produce thermal equilibrium), except that collective
repulsion could still support a core. Streaming motions would be required to explain the
extended parts of the halo outside the core, as with conventional noninteracting dark
matter.  Further discussion of these points occurs in \S5.



The paper is structured as follows.  In \S2, we compute the scattering cross section and pressure.  In \S3, we present isothermal spherical halo models.  In \S4, we combine the constraints of \S2 and \S3 and show that they cannot simultaneously be satisfied.  We discuss our results and conclude in \S5.  In two Appendices, we discuss the calculations behind the equation of state and examine the drag on a perturbation to the halo, e.g. a rotating galactic bar.

Throughout, $\sigma$ denotes a cross section and $m$ the dark matter particle's mass.  $\nu$ is always a number density. We define all other symbols where they are used, and also provide a table of the main symbols defined in this paper and their meanings for easy reference (Table 1).

\section{Scattering Cross Section and Pressure}
\subsection{Details of the RDM}

As preliminary to the scattering cross section and pressure, we present an overview of the
RDM model, beginning with the relativistic formalism, moving on to discuss the minimum
core size, and concluding with other remarks.  Like the axion, the DM particles are
bosonic and are supposed to be born in a Bose-Einstein condensate (BEC) in the early universe
(Goodman 2000, Peebles 2000).  Peebles and Goodman both argued that if present-day
repulsive dark matter derives from a relativistic scalar field, then it has acceptable
behavior in the early universe, \emph{i.e.} it does not suppress large-scale structure or
unduly affect primordial nucleosynthesis.

The dark matter particles' interaction would naturally be short-range, in fact point-like,
if it corresponded to the non-relativistic limit of a massive complex scalar field with
mass $m$ and a momentum-independent self-interaction term
$V(\phi,\phi^*)=\lambda(\phi^*\phi)^2$. 
We therefore treat the RDM as an assemblage of
non-relativistic point particles of mass $m$ having a two-body interaction potential $U$ whose
range is small compared to the particles' de Broglie wavelength, so that
$U(\bs{x}_1-\bs{x}_2)\to U_0\delta(\bs{x}_1-\bs{x}_2)$.  The constant $U_0\equiv \tilde
U(0)$ is the Fourier transform of $U(\Delta\bs{x})$ evaluated at zero momentum.

At tree level, $U_0=\lambda\hbar^3/4m^2c$ in conventional units, $\lambda$ being
dimensionless and nonnegative.  In contrast to Goodman (2000), we have taken the field to be complex so that
the particles are conserved, being protected from mutual annihilation via this
same interaction term by a global phase symmetry $\phi\to e^{i\psi}\phi$.\footnote{What is
  actually conserved is the (non-electromagnetic) ``charge'' represented by the number of
  particles minus the number of antiparticles.  For a
  relativistic chemical potential $\mu=\mu_{\textsc{nr}}+mc^2$, with nonrelativistic counterpart
  $\mu_{\textsc{nr}}$, in the nonrelativistic limit the antiparticles become so sparse that
  ``charge'' can be identified with particle number.}  Otherwise the particles would annihilate via a
cascade of stimulated emission in less than a Hubble time, unless their mass were so small
and their interaction so weak as to preclude the establishment of local thermodynamic
equilibrium (Riotto \& Tkachev 2000).

In the condensate, pressure varies only with density.  With a quartic self-interaction
potential such as we have adopted above, Goodman (2000) notes that in the non-relativistic
limit $P\ll\rho c^2$, pressure and density are related as in an $n=1$ Emden polytrope,
$P=K\rho^2$, with $K=U_0/2m^2$.  The density profile is
\begin{equation}\label{eq:emdenrho}
\rho(r)=\rho(0)\frac{\sin(\pi r/\rc)}{\pi r/\rc}
\end{equation}
with $0\le r\le \rc$ and core radius
\begin{equation} \label{eq:rc}
\rc=\sqrt{\frac{\pi K}{2 G}}=\sqrt{\frac{\pi U_0}{4 Gm^2}}\,.
\end{equation}
The total mass within $\rc$ associated with the profile \eqref{eq:emdenrho} is $M_{\rm c}
= 4\rho(0)\rc^3/\pi$.   The velocity $\vc$ of a circular orbit at the edge of the core is
therefore given by
\begin{equation}
  \label{eq:vc}
  \vc^2 \equiv \frac{GM_{\rm c}}{\rc} = \frac{4 G\rho(0)\rc^2}{\pi} = 2K\rho(0) =
  \frac{\rho(0) U_0}{m^2}\,,
\end{equation}
in which the second and third equalities result from eliminating $\rc^2$ using
eq.~\eqref{eq:rc}. 

If all of the DM were still in the condensate, then all non-rotating dark halos in virial
equilibrium would have this size and density profile independent of their masses.  Clearly
this would not be acceptable, especially if the core radius $\rc\sim 1\mbox{ kpc}$ as
required to match the DM cores of dwarf galaxies.  However, more complex and extended profiles can
be obtained if the DM has a finite temperature, as we discuss in the remainder of this paper, or where it has not
yet reached thermal or even hydrostatic equilibrium.

As we will show, the scattering cross section per unit mass is
$\sigma_{\rm scatt}/m = mU_0^2/2\pi\hbar^4$.  Since this involves a combination of $m$ and
$U_0$ independent from that appearing in the core radius \eqref{eq:rc}, the degree of collisionality of
RDM is independent of its minimum core size.  Thus it seems possible to evade the
constraint on $\sigma_{\rm scatt}/m$ set by the Bullet Cluster (Randall et al. 2008).
However, as will be shown, one then encounters difficulties with halo rotation curves.


\subsection{Scattering Cross Section}\label{subsec:scattering}
In the first-order Born approximation, the interaction potential
$U(\bs{x}_1,\bs{x}_2)=U_0\delta(\bs{x}_1-\bs{x}_2)$ between identical nonrelativistic
bosons entails the scattering cross section
\begin{equation}
\sigma_{\rm scatt}=\frac{m^2U_0^2}{2\pi\hbar^4}.
\label{e:scattcrosssxn}
\end{equation}
We can eliminate $U_0$ and rewrite $\sigma_{\rm scatt}$ in terms of the core radius \eqref{eq:rc}:
\begin{equation}
\sigma_{\rm scatt}=\frac{8G^2m^6\rc^4}{\pi^3\hbar^4}.
\label{e:scattrc}
\end{equation}
From the Bullet Cluster, we have the constraint that $\sigma_{\rm scatt}/m<1.25\;\rm cm^2\;\rm  g^{-1}$ (Randall et al. 2008); substituting eq. (\ref{e:scattrc}) into this relation implies that
\begin{equation}
m<9.6\times10^{-4}\times\left(\frac{\rc}{1\;\rm kpc}\right)^{-4/5}\;\rm eV/c^2.
\label{e:const1}
\end{equation}
We emphasize that $\rc$ in eq.~\eqref{e:const1} is the minimum size of a halo core supported only by repulsion; a larger
core radius is allowed for a given particle mass if the core is partly or entirely supported by random motions, as it must be for non-interacting DM.

Now, the Bullet Cluster bound was derived from N-body simulations of classical particles with dynamics specified completely by gravity and contact collisions.  Furthermore, the initial halo density profiles before the merger were King profiles.  Since the DM we consider here is bosonic, and our initial density profile does differ from the King profile, it is worth pausing to establish that Randall et al.'s bound really does apply.  In both their simulations and our model, the probability of the $i^{th}$ particle's scattering is
\begin{equation}
P_i=\rho_i\sigma_{\rm scatt} v_{\rm{rel}}\Delta t.
\end{equation}
$\rho_i$ is the local density, $v_{\rm{rel}}$ the relative velocity between the $i^{th}$ particle and its nearest neighbor, and $\Delta t$ the time-step. 

First, consider the local density.  As mentioned, in Randall et al. the initial conditions are that the two merging components have King density profiles, $\rho(r)=\rho(0)\left[1+\left(r/r_{\rm{c}}\right)^2\right]^{-3/2}$, whereas our density profile is as in Figure 2.  Two points defuse this concern.  One, Randall et al. also simulate the collision of a King and a Hernquist density profile, leading them to claim their results are only weakly-dependent on the initial mass profiles.  Two, the average density in our core is $\sim.6$ that in the King core.  This suggests that the maximum allowed scattering cross section in our model would differ by an order unity factor from the bound quoted above at worst.  As we will show, this bound would have to be orders of magnitude greater than what it is to permit bosonic repulsive DM in thermal equilibrium.  Hence an order unity change in the bound would not alter our conclusion.

Now, consider the relative velocity, which in our model might be expected to differ from that in Randall et al.'s because in the core the DM is in a condensate governed by the Gross-Pitaevskii or non-linear Schr$\rm{\ddot{o}}$dinger equation.  However, as we note in \S5.3, there is a minimum DM particle mass for thermal equilibrium to have been reached by now. This minimum mass implies the maximum thermal de Broglie wavelength possible for the DM particle is $\sim.5\;\rm{km}$.  This is the scale on and below which quantum effects, such as the macroscopic shared wave-function of the particles, governed by the GP Equation, become significant.  Clearly, it is far smaller than what would be resolved by a simulation.  Hence, an N-body simulation of the DM we consider here would not differ in this respect from Randall et al.'s---the GP Equation's dynamics would simply not be important on a cosmological scale, except as encoded in the halo density profile, an issue already dealt with above.

Still on the subject of the relative velocity, it has been pointed out that two interpenetrating streams of pure RDM condensate will not
scatter from one another and dissipate if their relative velocity is less than $v_{\rm
  crit} =\sqrt{2U_0\rho/m^2}$ for the same reasons as in a conventional superfluid
(Goodman 2000).  Despite this, the Bullet Cluster constraint still applies.  If $K$ is
chosen so that the minimum core radius of a dwarf galaxy is $\sim 1\,{\rm kpc}$,
then the critical velocity there will be
comparable to the circular velocity at the edge of the core,
$\vc\lesssim 100\,{\rm km\,s^{-1}}$.  From these values of $\vc$ and $\rc$, one infers a typical dark-matter density
$\rho_{\rm c}\sim 20\,m_H\,{\rm cm^{-3}}$ ($\approx 0.5\,{\rm M_\odot\,pc^{-3}}$).
The critical velocity in clusters will be lower because of the lower mass
density, $\lesssim 1\,m_H\,{\rm cm^{-3}}$.  

Now, Randall et al. estimate $v_{\rm{rel}}$ for the
Bullet Cluster as $4700\,{\rm km\, s^{-1}}$, while Springel and Farrar (2007) estimate
$2860\,{\rm km\, s^{-1}}$.  Either is much larger than $v_{\rm{crit}}$.  Hence there is no
question that the RDM particles will scatter each other, rather than the two merging components frictionlessly interpenetrating each other as would be the case were $v_{\rm{rel}}< v_{\rm{crit}}$.  Indeed, for $v_{\rm rel}>v_{\rm crit}$, as it is here, it is likely that the coherent state of the particles would be destroyed; the DM would no longer be in a condensate.  Nonetheless, subsequent to the merger it might cool to reach gravitational equilibrium and in the process re-condense.  

Finally, our discussion would be incomplete without acknowledging several other upper bounds on $\sigma/m$.  Brada\u{c} et al. (2008) use the merging cluster MACS J0025.4--1222 in similar fashion to the Bullet Cluster and obtain the order-of-magnitude estimate $\sigma/m< 4\;\rm{cm}^2\;\rm{g}^{-1}$.  It is worth noting that even were this looser limit used, it would still be too low to permit bosonic repulsive DM in thermal equilibrium.  Other more stringent limits are available, however. Miralda-Escud\'e (2002) combines ellipticity measurements with the fact that DM collisions isotropize the DM's stress-energy tensor and hence lead to more spherical galaxies.  He finds $\sigma/m< .018\;\rm{cm}^2\;\rm{g}^{-1}$.  As we discuss in \S5.2, $\sigma/m$ must be  $\sim 1\;\rm{cm}^2\;\rm{g}^{-1}$ to allow local and global thermodynamic equilibrium, which our work will require.  Hence, if Miralda-Escud\'e's bound truly holds, it is an independent reason to doubt that DM can be in thermal equilibrium. Lin et al. (2011) present another bound also using the idea that only limited isotropization is observationally acceptable; it is $\sigma/m < .0025\;\rm{cm}^2\;\rm{g}^{-1}$.  This bound too offers reason to doubt that DM can be in thermal equilibrium.  

Nonetheless, given the uncertainties associated with limits on DM's self-interaction cross section, in this paper we have chosen to adopt the view that repulsive, bosonic DM in thermal equilibrium may still be possible, and must be more fully considered before it can be decisively judged.

\subsection{Pressure}\label{subsec:pressure}
Our main conclusions depend upon certain expected properties of the finite-temperature
equation of state (hereafter EOS).  Before presenting mathematical details, we address the
general physical regime that we expect for this hypothetical RDM gas.

The limits of \S\ref{subsec:scattering} imply that the interaction energy between any
single pair of particles is $\ll m \vc^2$, $\vc$ being a characteristic virial or
circular velocity.  However, the \emph{total} energy of interaction between a given particle and
all of its neighbors is $\sim m \vc^2$ in the core where repulsion
balances gravity.  To match observed rotation curves, the halo should have an extended,
approximately power-law envelope where the density is much lower than in the core.  If the
collisionality, though weak, suffices to establish local thermal and hydrostatic
equilibrium (\S\ref{subsec:thermeq}), the envelope must be supported mainly by microscopic
thermal motions rather than the interparticle repulsion since the latter scales with the
density.  Hence the temperature in the envelope must be virial, $\kB T\sim m \vc^2$.  We
take the temperature of the gas to be the same in the halo as in the envelope.  This is
done partly for simplicity, but the limits above imply that the collisional mean-free path
is comparable to the size of the galaxy, if not larger, so thermal conduction should be
efficient.  Finally, the repulsive interaction has a very short range, because we assume
that it derives from a momentum-independent quartic potential of a complex scalar field.
This range is much smaller than the thermal de Broglie wavelength
\begin{equation}
\Lambda_{\rm dB}\equiv\frac{h}{\sqrt{2\pi m k_{\rm B}  T}}\,,
\end{equation}
which in turn is much smaller than galactic scales unless the
particle is extremely light.

With these assumptions, the RDM should behave as a nearly ideal boson gas, a type of
system that has been studied extensively in connection with superfluidity and
Bose-Einstein condensates.  At least to first order in perturbation theory, the short-range
two-body repulsion can be represented by a contact potential
$U_0\delta^{(3)}(\bs{r}_1-\bs{r}_2)$ described by the single parameter $U_0$, or equivalently,
the ``scattering length''
\begin{equation}
l=\frac{mU_0}{4\pi\hbar^2}.
\label{e:ldef}
\end{equation}
This is not to be confused with the collisional mean-free path, which depends on density.
In the first-Born approximation, $l=(\sigma_{\rm scatt}/8\pi)^{1/2}$.
The weakness of the
individual pairwise interactions is expressed by $l\ll\Lambda_{\rm dB}$.

Even with the idealization of a contact potential, the thermodynamics of a boson gas has
not been solved exactly.  Proukakis \& Jackson (2008, hereafter PJ) review many of the
approximations that have been used.  The one we have adopted is what PJ call
``Hartree-Fock'' (HF).  In this approximation, the gas is described by the occupation numbers
of single-particle states---in our case, plane-wave momentum states, since
$\Lambda_{\rm dB}$ is much smaller than scales on which the density or pressure varies.  The
grand-canonical partition function and pressure are derived semiclassically along lines
similar to textbook derivations for a non-interacting boson gas.  Some
tricks are used to incorporate the interactions, as detailed in Appendix A.  

At a given temperature $T$ and particle mass $m$, the critical density above which the
condensate appears in a non-interacting gas is\footnote{We use $\nu$ for number per unit volume
and reserve $n$ for mode occupation number.}
\begin{equation}
\nu_{\rm crit}=\zeta\left(\frac{3}{2}\right)\Lambda_{\rm dB}^{-3}\approx 2.6\,\Lambda_{\rm dB}^{-3}\,,
\label{eq:nucrit}
\end{equation}
i.e. approximately one particle per cubic de Broglie wavelength.  In our HF approximation,
the critical density for the condensate is also given by eq.~\eqref{eq:nucrit}.

The HF equation of state behaves as expected in highly degenerate and dilute limits,
respectively $\nu\gg\nu_{\rm crit}$ and $\nu\ll\nu_{\rm crit}$.  In the dilute limit, one
recovers the pressure of an ideal gas because the repulsion becomes negligible.  In the
opposite limit, the HF approximation gives the polytropic equation of state $P=K\rho^2$,
as does every approximation that is consistent with the Gross-Pitaevskii equation
\eqref{eq:NLSE}.  

The HF approximation is least accurate in an intermediate regime where the condensate is
present and the repulsive and thermal energies per excitation are comparable, $\nu
U_0\sim\kB T$ (see Pethick \& Smith 2002).  In such cases, the thermal excitations are not well described by single
particles but rather by collective oscillations---quasiparticles.  The dispersion relation for
these excitations is eq.~\eqref{eq:dispersion}, which follows from linearisation of the
Gross-Pitaevskii equation \eqref{eq:NLSE}.  It can be shown that our HF
approximation effectively replaces eq.~\eqref{eq:dispersion} with (neglecting the
gravitational term)
\begin{equation}  \label{eq:HFdispersion}
  \hbar\omega_{k}=\nu U_0 + \frac{(\hbar k)^2}{2m}.
\end{equation}
If this were correct, there would be a minimum energy for the excitations (viz. $\nu U_0$)
even at zero momentum, whereas the more correct relation \eqref{eq:dispersion} shows that
the excitations with wavenumber $k\ll 2 m \cs/\hbar\equiv k_{\rm s}$ behave as
phonons---quantized sound waves---with effective sound speed $\cs=\sqrt{\nu U_0/m}$.
These phonons are not conserved: they are excited and damped by 3-mode
nonlinear interactions that are not part of our HF approximation (which does however
incorporate  particle-conserving 4-mode interactions).

In order to assess the consequences of neglecting the quasiparticles for our equation of
state, it is useful to introduce a dimensionless parameter that compares the repulsive and
thermal energies at the critical density:
\begin{equation}
\theta=\frac{U_0\nu_{\rm crit}}{k_{\rm B} T}= 2\zeta(\tfrac{3}{2})\frac{l}{\Lambda_{\rm dB}}\,.
\label{e:theta}
\end{equation}
The limits on collisionality from the Bullet Cluster imply that
$\theta\ll 1$; in fact, using eqs.~\eqref{eq:rc}, \eqref{e:scattrc}, and
\eqref{eq:isosphere} to eliminate $U_0$, $m$, and $T$ in favor of $\rc$, $\vci$ (the
asymptotic halo velocity) and $\sigma_{\rm scatt}/m$, one finds
\begin{align}
  \label{eq:themax}
  \theta&=  2^{-12/5}\pi^{-1/10}\zeta(\tfrac{3}{2})
\left[\frac{\hbar\vci^5}{G^3\rc^6}\left(\frac{\sigma_{\rm scatt}}{m}\right)^4\right]^{1/5}\nonumber\\
&\approx 7\times 10^{-21}\times\left(\frac{\vci}{100\,{\rm km\,s^{-1}}}\right)
\left(\frac{\rc}{1\,{\rm kpc}}\right)^{-6/5}\left(\frac{\sigma_{\rm scatt}/m}{1.25\,{\rm cm^2\,g^{-1}}}\right)^{4/5}
\end{align}
An independent reason to expect $\theta$ to be small is the assumption that RDM is the
nonrelativistic limit of a charged scalar field with interaction term $\lambda|\phi|^4$ (see Kapusta \& Gale 2006 for discussion of this formalism):
if $U_0 = \lambda\hbar^3/4m^2c$, then
\begin{equation}
  \label{eq:theta_from_lambda}
  \theta = \lambda\frac{\zeta(3/2)}{16\pi^{3/2}} \left(\frac{\vci}{c}\right) \approx 10^{-5}\times
\left(\frac{\vci}{100\,{\rm km\,s^{-1}}}\right)\,\lambda\,,
\end{equation}
and $\lambda$ must be less than unity in order that perturbation theory make sense for the
relativistic theory.

Let us compare the maximum energy of a phonon-like excitation, $\epsilon_{\rm s}= \hbar
\cs k_{\rm s}= m\cs^2$, with the typical thermal excitation energy, $\kB T$.  We find from
the correct dispersion relation \eqref{eq:dispersion} that thermal excitations will be
more like free particles (i.e. $\epsilon\approx (\hbar k)^2/2m$) than phonons unless
$\nu\gtrsim \theta^{-1}\nu_{\rm crit}$.  

At high densities satisfying the latter condition,
the phonon pressure has the Debye form
\begin{equation}
  \label{eq:qp_pressure}
  P_{\rm ph}\approx \frac{\pi^2}{90}\frac{(\kB T)^4}{(\hbar \cs)^3}\,.
\end{equation}
We can compare this with the pressure of the
unexcited condensate, $P_0= U_0\nu^2/2$. 
One finds that $P_{\rm ph}\le P_0$ when $\nu\gtrsim
2.0\,\theta^{-5/7}\nu_{\rm crit}$.\footnote{The numerical constant is actually
  $[8\pi^{11}/(45\zeta(3/2))^2]^{1/7}$.}  For $\theta\ll 1$, this minimum is lower
than the density at which eq.~\eqref{eq:qp_pressure} is valid.  That means that for any density for which the thermal excitations are more like phonons than free particles (and the latter are already dealt with in our HF approximation), the pressure due to them will be less than the pressure of the unexcited condensate. In other words, the
\emph{thermal} contribution to the pressure is dominated by quasiparticles rather than
``free'' particles only when the density is so high that the total pressure is well-approximated by
the zero-temperature relation $P=K\rho^2$ anyway.

In summary, since $\theta$ must be small to satisfy the Bullet Cluster constraint (see eq. (\ref{eq:theta_from_lambda})), we may
neglect the quasiparticles and use the HF approximation of Appendix A to calculate the
pressure as a function of density and temperature in RDM halos.

After these preliminaries, we now present our approximate equation of state in terms of a
scaled pressure $\hat{P}=P/\nu_{\rm crit}k_{\rm B}T$, a scaled total number density
$\hat{\nu}=\nu/\nu_{\rm crit}$, and a scaled number density in the condensate
$\hat{\nu}_0=\nu_0/\nu_{\rm crit}$.
In terms of these parameters, the equation of state is
\begin{equation}
\hat{P}=\left(\hat{\nu}^2-\frac{1}{2}\hat{\nu}_0^2\right)\theta+\frac{\mathrm{Li}_{5/2}(z)}{\zeta\left(3/2\right)},
\label{e:eos}
\end{equation}
with
\begin{equation}
\hat{\nu}=\frac{\mathrm{Li}_{3/2}(z)}{\zeta\left(3/2\right)}+H(\hat{\nu}_0)
\label{e:hatnu}
\end{equation}
and
\begin{equation}
z=\exp[-\theta\hat{\nu}_0],\:\hat{\nu}_0>0.
\label{e:zofnu0}
\end{equation}
$H$ is the Heaviside function ($0$ for $\hat{\nu}_0<0$, $1$ for $\hat{\nu}_0>0$).  When
$\hat{\nu}_0=0$, $z$ is defined implicitly by eq. (\ref{e:hatnu}).

\begin{figure}
\includegraphics[scale=.4]{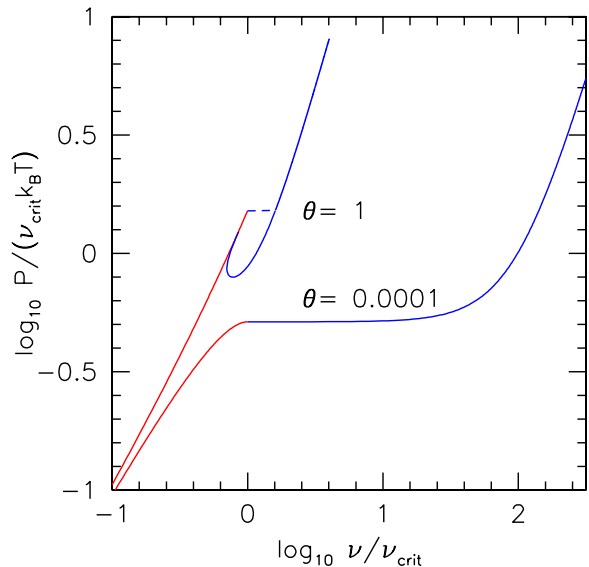}
\centering{}\caption{A log-log plot of scaled pressure $\hat{P}\equiv P/\nu_{\rm crit}k_{\rm B}T$ versus scaled number density $\hat{\nu}\equiv \nu/\nu_{\rm crit}$ for interaction strengths $\theta=1$ and $\theta =10^{-4}$ (eq. (\ref{e:eos})). The red curve is when no condensate is present, and the blue curve (only that to the right of and above the dashed line for $\theta=1$) is when the condensate is present.  For $\theta=1$, the loop below the dashed line is unphysical: as the pressure increases beginning around $\log_{10}\hat{P}=.2$, the density experiences a large jump.  For $\theta=10^{-4}$, the pressure barely varies over a large range in density; this leads to the sudden drop in the density profile seen in Figure~\ref{f:densityprof}. }
 \vspace{0.2in}
\label{f:pvsnu}
\end{figure}

A log-log plot of the equation of state (eq. (\ref{e:eos})) is shown in Figure~\ref{f:pvsnu} for
$\theta=1$, corresponding to strong scattering (although the steps leading to eq. 
(\ref{e:eos}) cannot be justified unless $\theta\ll 1$) and for $\theta=10^{-4}$,
corresponding to moderately weak scattering.  The condensate is absent along the red
section, and present along the blue.  The part of the curve beneath the dashed line
segment is unphysical; it can be shown that the integral of $\hat P\ud\hat\nu$ vanishes
when taken around the loop defined by the dashed segment and the unphysical lobe below it.
The endpoints of the dashed segment define two phases in contact, and the density jump at
the phase transition scales $\propto\theta$ when $\theta\ll1$.
The actual phase transition is rather weak in the weakly interacting regime
($\theta\ll 1$) where our equation of state is valid.  

We know from the discussion above that this equation of state is only approximate.  What
matters for the computation of halo rotation curves, however, is the following property
that the foregoing assures us is shared by the exact EOS:
when $\theta\ll 1$, the pressure supplied by the condensate is very small at the critical
density compared to the pressure supplied by the non-degenerate component, and the latter
is approximately that of a non-interacting gas.  Therefore, as the pressure increases by a
modest factor above its value at $\nu_{\rm crit}$---as it doubles, for example---the density
must increase by a large factor $\sim\theta^{-1/2}$.
This property of the EOS, together with the assumption of hydrostatic equilibrium,
leads to a large jump in dark-matter density near the edge of the degenerate core.

\section{Density Profiles for an Isothermal Spherical Halo}
\begin{figure}
\includegraphics[scale=.4]{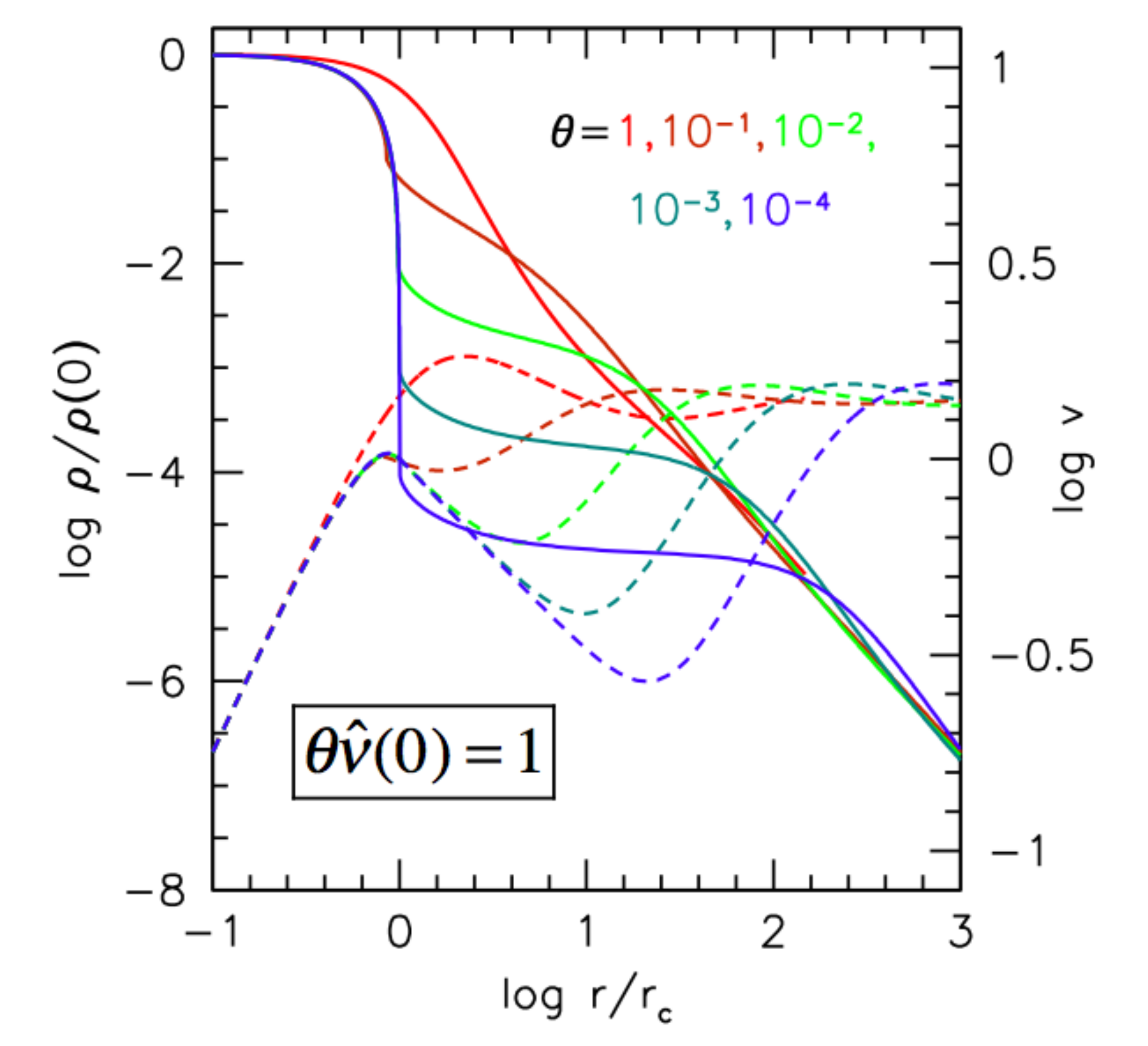}
\centering{}\caption{Self-gravitating isothermal spheres of RDM for various interaction
  strengths $\theta$ (eq. (\ref{e:theta})). Solid curves and left axis: log-log mass
  density. Dashed curves and right-hand axis: log-log rotation curves. Notice the severe
  drops in the density profile for $\theta \ll 1$; these come from the small change in
  pressure over a large range in density we point out in Figure 1, and lead to the
  constraint that $\theta \geq10^{-4}$ for realistic rotation curves.}
 \vspace{0.2in}
\label{f:densityprof}
\end{figure}

Using the equation of state, we solve the equations of hydrostatic equilibrium
\begin{equation}
\frac{dP}{dr}=-\frac{GM_r\rho}{r^2}
\end{equation}
and
\begin{equation}
\frac{dM_r}{dr}=4\pi\rho r^2
\end{equation}
The isothermal spherical halo we obtain thereby is the simplest model for a dark matter
halo in virial equilibrium that has the correct flat rotation curve asymptotically.
At large radii where the density is sufficiently small, the pressure $P\to\nu\kB T=\rho\kB
T/m$, so that the solution tends to a classical isothermal sphere:
\begin{align}
  \label{eq:isosphere}
  \rho(r) &\to \frac{\kB T}{2\pi Gm}r^{-2}\,,&\mbox{and}&\nonumber\\
\frac{GM_r}{r} &\to \frac{2\kB T}{m} \equiv \vci^2  & \mbox{as } r&\to\infty\,.
\end{align}
This will provide an asymptotically flat rotation curve with amplitude $v_{\rm
  c,\infty}$.  Within the core ($r<\rc$), the condensate dominates, and the density
profile is similar to that for an $n=1$ Emden polytrope. This is unsurprising because for
a pure condensate the equation of state $is$ just that for an $n=1$ polytrope.  The
inferred contribution of dark halos to the rotation curves of observed galaxies leads us
to expect that the asymptotic velocity $\vci$ should be within a factor $\sim
2$ of the circular velocity $\vc$ at the edge of the core (eq.~\eqref{eq:vc}).  This
connects the temperature---or more precisely, the ratio $T/m$---to the sound speed at the
center of the degenerate core, which depends in our model only upon the zero-temperature
limit of the equation of state: $\cs^2(0)=2K\rho(0)$.  At intermediate radii however, as
can be seen from Figure~\ref{f:densityprof}, the density profile and circular velocity are
quite different from the classical isothermal sphere and depend strongly on the
interaction parameter $\theta$. 

All models shown are scaled to the same central density, core radius and asymptotic
rotation velocity or velocity dispersion; this demands that the scaled central number
density $\hat{\nu}(0)=C \theta^{-1}$, $C$ a constant.  The reason for this is as follows.
The physical central density $\rho(0)=m\nu_{\rm{crit}}\hat{\nu}(0)$.  Eliminating
$\nu_{\rm crit}$ in favor of $\theta$ via eq.~\eqref{e:theta}, $U_0$ in favor of $\rc^2$
via eq.~\eqref{eq:rc}, and finally $\kB T/m$ via eq.~\eqref{eq:isosphere} yields $\rho(0)=(\pi
\vci^2/8G\rc^2)C$. On the other hand, $\rho(0)=(\pi \vc^2/4G\rc^2)$ from
eq.~\eqref{eq:vc}.  Thus the asymptotic circular velocity is related to the circular
velocity at the edge of the core by $\vci = \vc\sqrt{2/C}$.  Typical dwarf-galaxy rotation
profiles suggest that $\vci$ should be modestly greater than $\vc$, so we set $C=1$.  We
thus have a family of density profiles that depends only on $\theta$, the interaction
strength of the RDM, as demonstrated by Figure~\ref{f:densityprof}.

Perhaps surprisingly, it is the most strongly interacting model ($\theta=1$) that most
closely resembles the classical case.  At $\theta \ll1$, the density drops sharply outside
the core, by a factor $\sim\theta$.  It is then nearly constant out to $r\approx
\rc/\sqrt{\theta}$.  This occurs because, as discussed in \S\ref{subsec:pressure}, the
pressure of the condensate is proportional to $\theta(\nu/\nu_{\rm crit})^2$; the
condensate would not contribute to the pressure at all if $\theta=0$ (a noninteracting
gas).  In the limit $\theta=0$, the pressure of the non-degenerate component, and
therefore the total pressure, is independent of density once $\nu>\nu_{\rm crit}$;
increases in density merely increase the fraction of the particles in the condensate.  If
$\theta$ is small, the density must increase above $\nu_{\rm crit}$ by a factor $\sim
\theta^{-1/2}$ in order to double the pressure.  Hence in hydrostatic equilibrium, a very
large increase in density must occur over a small range in radius at the edge of the
region where the condensate exists.  The structure of the low-$\theta$ halos in
Fig.~\ref{f:densityprof} is reminiscent of a red giant, where the central parts are
supported by degeneracy pressure and there is a distended envelope due to the large
entropy increase across the hydrogen-burning shell.

This density drop is a key feature of our model because it leads to a dip in the rotation
curve at the edge of the core.  For $\theta=10^{-4}$, the velocity dips by a factor $\sim
2$, and for smaller values of $\theta$ the dip is even more pronounced.  Such features
appear to conflict with the inferred contribution of dark matter to galactic rotation
curves.  Therefore, if RDM halos really were isothermal, we would surely require $\theta\geq
10^{-4}$.

Requiring $\theta\geq10^{-4}$ leads to a lower bound on the mass of the RDM particle.\footnote{Kouvaris \& Tinyakov (2011) provide a lower limit on the mass and cross section of bosonic self-interacting DM using neutron star observations, but theirs is not in conflict with the Bullet Cluster upper bound.  Hence our more stringent bound from the rotation curves is needed to see that the DM in thermal equilibrium we consider here is untenable.}
Replacing $(2\kB T/m)^{1/2}$ with the asymptotic circular velocity \eqref{eq:isosphere}
yields $\Lambda_{\rm dB}\approx\sqrt{4\pi}\hbar/m\vci$, which we insert into the definition
\eqref{e:theta} of $\theta$.  We relate $l$ in that equation to $U_0$ via
eq.~\eqref{e:ldef} and $U_0$ to $\rc$ via eq.~\eqref{eq:rc}.  These manipulations yield
\begin{equation}
\theta\approx \zeta(\tfrac{3}{2})\frac {Gm^4\rc^2\vci}{\pi^{5/2} \hbar^3}.
\label{e:thetatwo}
\end{equation}
Requiring that $\theta\geq10^{-4}$ means
\begin{equation}
m\geq 10.\times\left(\frac{\vci}{100\;\rm km\;\rm s^{-1}}\right)^{-1/4}\left(\frac{\rc}{1\;\rm kpc}\right)^{-1/2}\;\rm eV/c^2.
\label{e:const2}
\end{equation}
\section{Ruling out RDM in Thermal Equilibrium} 
Consider jointly eqs. (\ref{e:const1}) and (\ref{e:const2}).  As $\rc$ increases, the lower bound on $m$ will  fall, but evidently so will the upper bound, so we cannot hope to leave $\vc$ fixed and tune $\rc$ such that both constraints are satisfied.  Could we leave $\rc$ fixed and raise $\vc$ enough so that eq. (\ref{e:const2}) aligns with eq. (\ref{e:const1})?  Certainly not; since eq. (\ref{e:const2}) depends so weakly on $\vc$, to achieve this would demand $\vc\simeq10^{16}\; {\rm km}\; {\rm s}^{-1}$, considerably in excess of the speed of light! Figure 3 shows the ratio $\mu\equiv {\rm max}(m)/{\rm min}(m)$; for a viable dark matter particle it has to be greater than or equal to unity, which it clearly is not over the realistic parameter space for $\vc$ and $\rc$.

\begin{figure}
\includegraphics[scale=.75]{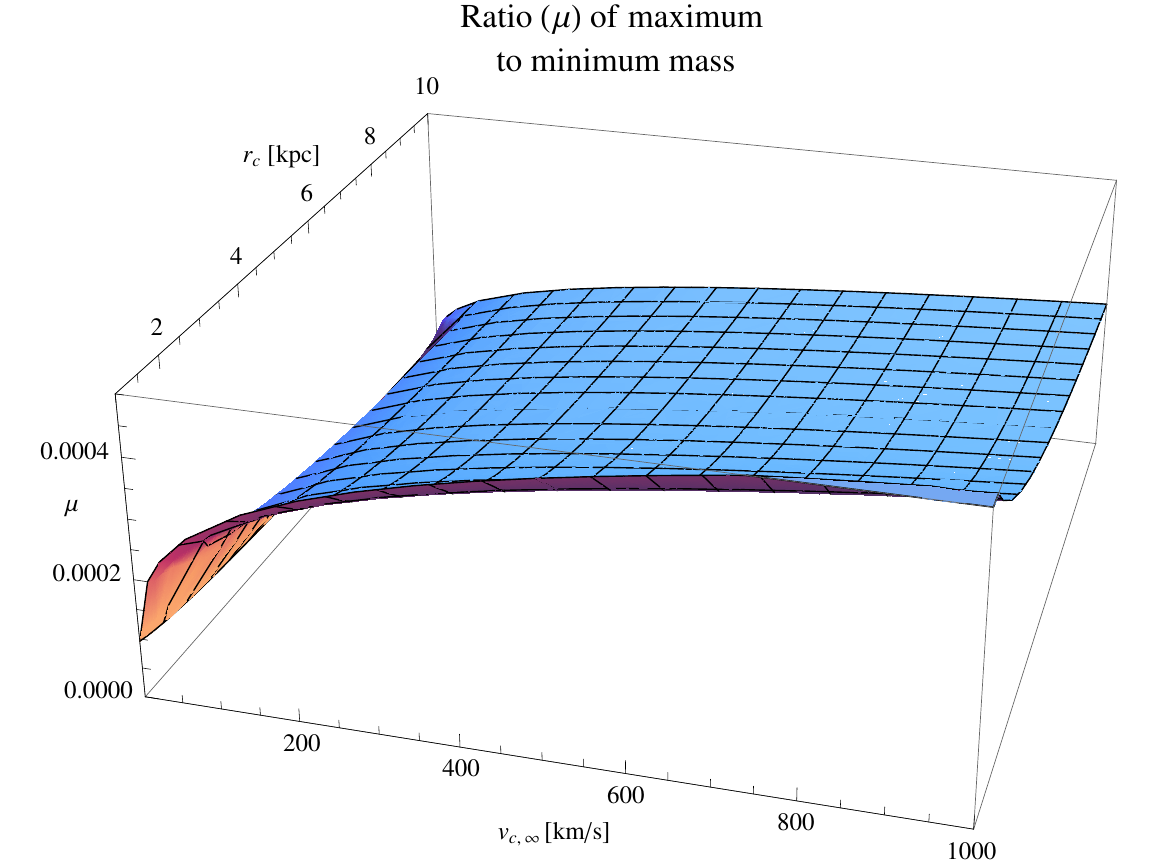}
\centering{}\caption{Plot of ratio $\mu$ of maximum to minimum mass from two constraints eqs. (\ref{e:const1}) and (\ref{e:const2}); horizontal axes are $\vc$ in $km/s$ (front) and $\rc$ in $kpc$ (left-hand side).  For a viable model, clearly the maximum mass must be greater than the minimum mass, so we would require $\mu\geq1$.  This is not achievable over the realistic parameter space for $\vc$ and $\rc$. }
\vspace{0.2in}
\label{f:constplot}
\end{figure}

\begin{table}
 \vspace{0.2in}
\caption{Symbols used in this work}
 \vspace{0.04in}
\centering

    \begin{tabular}{ | l | l  p{1.2 cm} |}
    \hline
    $\nu$&number~density&\\ \hline
    $\hat{\nu}$ &scaled~number~density&\\ \hline
    $\nu_0$&condensate~number~density&\\ \hline
    $\nu_{\rm crit}$&density~where~cond.~appears&\\ \hline
     $\theta$&interaction~parameter&\\ \hline
    $r_{\rm c}$&halo~core~radius&\\ \hline
    $\vc$&halo~circular~velocity&\\ \hline
    $U_0$&Fourier~transform~of~pot'ial~at~$\vec{p}=0$&\\ \hline
    $m$&RDM~particle~mass&\\ \hline
    $ \sigma_{\rm scatt}$&RDM~particle~scattering~cross~sxn.&\\ \hline
    $l$&scattering~length&\\ \hline
    $\Lambda_{\rm dB}$&RDM~de~Broglie~wavelength&\\ \hline
  
    \end{tabular}
     \vspace{0.2in}
     \label{t:Boss}
\end{table}

\section{Discussion and Conclusions}

\subsection{Context for our work}
As we note in the Introduction, self-interacting, bosonic dark matter models have been considered before.  For a detailed review of the history of work on astrophysical-scale bosonic objects, e.g. boson stars (first proposed by Kaup (1968) and Ruffini \& Bonazzola (1969)), as well as galactic halos (first suggested by Baldeschi et al. (1983)), we refer the reader to  Chavanis 2011.  Here, we review what is necessary to contextualize our own work.

As Chavanis notes, Baldeschi et al. considered the possibility that DM halos might be
composed of a non-degenerate component and a condensate.  However, in this model and
subsequent studies (e.g. Sin 1994), the possibility of self-interaction was ignored.  Many
scalar field models (Schunck 1998, Matos \& Guzman 1999, see Chavanis 2011 for further
discussion) were proposed that also assumed no self-interaction.  These models all
required small masses ($m\sim10^{-24}\;{\rm eV/c^2}$) to achieve a sufficiently large core
radius in agreement with observation.  
To avoid this ``unnaturally small mass,'' self-interaction was proposed: see Colpi et al. 1986, Lee \& Koh 1996, and Arbey et al. 2003, as well as the references mentioned in the Introduction.  

Goodman (2000) pointed out that this would lead to a minimum halo core size $r_{\rm
  c}=3\sqrt{U_0/4\pi G m^2}$, and Lee \& Lim (2010) show that there is also a minimum
mass, $M_{\rm min}\simeq \hbar^2/G\lambda_{\rm c}m^2$, with $\lambda_{\rm c}$ the DM
particle's Compton wavelength. Rindler-Daller \& Shapiro (2010) observe that vortices
should form for a strongly self-interacting condensate, and that this may affect the
density profile (2011).  Bohmer \& Harko (2007) pick up where Goodman left off and
calculate rotation curves for a repulsive, completely Bose-Einstein condensed halo, noting
that the gravitational lensing prediction differs from CDM.  However, as Chavanis
comments, they ignore the quantum pressure, i.e. the additional pressure due to Heisenberg
uncertainty. Chavanis therefore connects the non-interacting (but including quantum
pressure) limit of Ruffini \& Bonazzola (1969) with the interacting (but no quantum
pressure) limit studied by Bohmer \& Harko (2007), taking into account both the quantum
pressure and a repulsive (or attractive) self-interaction.\footnote{We neglect the quantum
  pressure because for it to be significant would require a mass so small as to have a
  Compton wavelength comparable to the core radius ($m\sim h/\rc\vc\sim 10^{-22}\,{\rm
    eV}$), as with Hu et al.'s (2000) ``fuzzy dark matter.''  This would be the type of
  ``unnaturally small mass'' that self-interaction was originally introduced to avoid.}
Chavanis obtains both analytical (paper I) and numerical (paper II) mass profiles and
rotation curves.  However, his work does not treat the case where the DM has a non-zero
temperature and there are both condensed and non-condensed components present.  This is
the gap our paper fills.

We would note that it is an important gap because as it turns out, the presence of a non-condensed component is what leads to the abrupt density drop in the halo profile (see Figure \ref{f:densityprof}).  This abrupt drop in turn leads to unrealistic rotation curves and so places a lower bound on the collisionality of the RDM (lower collisionality implies a steeper density drop).  It is the combination of this constraint with that the Bullet Cluster places on the scattering cross section that allows us to rule out repulsive, bosonic dark matter in thermal equilibrium.  Hence we can conclude that the effects of non-zero temperature are in fact critical to understanding the behavior of a bosonic, repulsive DM model in thermal equilibrium.

\subsection{Thermal Equilibrium}\label{subsec:thermeq}
Before closing, there are four issues we should briefly address.  First, as we noted in the Introduction, the conclusion we have presented can be avoided if the DM has not reached thermal equilibrium either locally or globally.  For our equation of state to be valid, we require local thermodynamic equilibrium, reached via collisions on a timescale $t_{\rm coll}\sim \lambda_{\rm mfp}/\vc$.  For our isothermal halo models to be valid, we also require global thermodynamic equilibrium, reached via conduction on a timescale $t_{\rm cond}\sim r^2/\vc\lambda_{\rm mfp}$, where $r$ is the radius of the DM halo and we estimate $r\sim r_{\rm c}$.  So unless $t_{\rm coll}$ and $t_{\rm cond}$ are both much less than the Hubble time $t_{\rm H_0}$, our halo models are invalid, in which case so is the constraint eq. (\ref{e:const2}), as it comes from demanding that the drop in the halo density profile not be too steep.  

If $\sigma_{\rm scatt}/m\approx 1\;{\rm cm^2\;g^{-1}}$ (near the upper end of the range allowed by the Bullet Cluster), then $r\sim \lambda_{\rm mfp}$ in the solar neighborhood, and $t_{\rm coll}\sim t_{\rm cond}\ll t_{\rm H_0}$ so we have both local and global thermodynamic equilibrium.  But if $\sigma_{\rm scatt}/m$ were significantly smaller than this value, the dark matter would not undergo sufficient scattering to reach local thermodynamic equilibrium (i.e. $t_{\rm coll} >\ t_{\rm H_0}$). This would invalidate our equation of state, and place us effectively in the limit of a non-interacting, axion-like particle.  There would be one salient difference from the non-interacting axion even in this regime, however: the collective repulsion could still provide a core, since as we have pointed out earlier, the collisionality and minimum core size are independent.  In short, in this limit, the upper bound eq. (\ref{e:const1}) on the mass still applies, but the lower bound eq. (\ref{e:const2}) does not.

\subsection{Supporting the degenerate core}
\label{subsec:thetasupport}
Second, we reiterate here that there is no conflict between a small interaction parameter $\theta$
and a large core, $\rc\gg\Lambda_{\rm dB}$, because the de Broglie wavelength, core
radius, and interaction parameter depend upon the particle mass $m$ and scattering length
$l$ in different combinations.  From eq.~\eqref{e:theta} we have that $\theta\sim
l/\Lambda_{\rm dB}$.  (To avoid distraction from the important scalings dimensionless
factors of order unity are suppressed here.)  The temperature $T$ of the hypothetical RDM
is not observable, but $\kB T/m\sim\vc^2$ \emph{is} observable, $\vc$ being a
characteristic circular velocity.  Thus $\Lambda_{\rm dB}\sim h/m\vc$ and $\theta\sim m l
\vc/h$.  On the other hand, eliminating $U_0$ from the core radius \eqref{eq:rc} in favor
of the scattering length \eqref{e:ldef} gives $\rc\sim (\hbar^2 l/Gm^3)^{1/2}$.  So at a
fixed value of $\vc$, we have $\Lambda_{\rm dB}\propto m^{-1}$, $\theta\propto m l$, and
$\rc\propto l^{1/2} m^{-3/2}$.  Eliminating the scattering length between these relations
and including the numerical factors produces eq. \eqref{e:thetatwo}, which clearly
shows that one can have an arbitrarily small $\theta$ at any given $\rc$ for sufficiently
small particle mass.  

If the mass were as small as $h/\rc\vc\sim 10^{-22}\,{\rm eV}/c^2$, then
the de Broglie length would be comparable to the core radius; for masses smaller than
this, the core would be supported by ``quantum pressure'' even without repulsive
interactions.  However, as eq. \eqref{e:const1} and the discussion in
\S\ref{subsec:thermeq} indicate, the scattering cross section is insufficient to establish
thermal equilibrium at galactic densities over a Hubble time if the particle mass is much
smaller than $10^{-3}\,{\rm eV/c^2}$.  Thus there is a wide range of masses---some 19
orders of magnitude---over which the cores of galaxies could be supported by repulsion
rather than a large de Broglie wavelength and
yet two-body collisions would be totally ineffective at thermalizing their distribution.
This regime, however, is not the main focus of the present paper.

\subsection{Vortices}
Third, we consider whether the vortex formation Rindler-Daller \& Shapiro (hereafter RDS) (2010) claim will occur in strongly self-interacting Bose-Einstein condensed DM would possibly change the rotation curve (RDS 2011) in a way to avoid the constraint we use here.


Some care is required.  RDS (2011) derive two constraints that both must be independently satisfied for vortices to form: a minimum mass, and a minimum interaction strength.  At first blush, we might take the minimum mass of eq. (\ref{e:const2}) and ask whether it exceeds their required minimum.  

However, this approach would be subtly flawed: the minimum mass of eq. (\ref{e:const2}) is based on the observationally disallowed density profile it would produce.  But this density profile assumes no vortices are present.  Hence, the minimum mass it implies is not the correct lower bound if vortices form.  Indeed, if vortices form, they would form only in the condensate core of the halo, and lower the density there while minimally affecting the non-degenerate envelope's density.\footnote{RDS's work is motivated by the fact that vortices form in BEC's produced in the laboratory (see e.g. Madison et al. 2000); there is no reason to expect them to form in the non-degenerate component of the halo.  The change in the density of the core would affect the gravitational potential experienced by the outer, non-degenerate envelope, but this effect would be at 2nd order.}  So at a given $\theta$, the ``cliff'' in the density would be smoothed out---thus allowing a lower minimum value of $\theta$ to be consistent with observation than if vortices are not considered.  Since $m\propto \theta^{1/4}$ from eq. (\ref{e:thetatwo}), this would lower the minimum allowed mass of eq. (\ref{e:const2}).

Indeed, it is worthwhile to ask just how much vortices could lower this minimum allowed mass of eq. (\ref{e:const2}).  In particular, could they lower it enough so that it became less than the maximum mass of eq. (\ref{e:const1}), thereby saving repulsive dark matter in thermal equilibrium from being ruled out by the mismatch of these constraints?  To get the constraints to match would require lowering the lower bound eq. (\ref{e:const2}) by a factor of $10^{-5}$, and hence $\theta$ by a factor of $10^{-20}$.  So the vortices would have to deplete the core density enough that a density profile corresponding to $\theta\simeq 10^{-24}$ was observationally allowed.  

Even with the mass lower by a factor of $10^{-5}$, $\Lambda_{\rm{dB}}\ll r_{\rm{c}}$, so
the contribution of quantum pressure to supporting the core would still be
negligible. Thus the core would still need to be supported by repulsion alone.  But $\theta\geq 10^{-9}$ is required to
fulfill this condition. So if the vortices could deplete the core's density sufficiently
to make our lower bound on $m$ consistent with our upper bound, we would lose the
degenerate core, the feature that motivated considering repulsive bosonic DM in the first
place!  In short, then, the presence of vortices will not affect our conclusion that
repulsive bosonic DM in thermal equilibrium is observationally ruled out.

For the sake of completeness, we now deal with the question of whether vortices will
indeed form.  The minimum mass for which our model can be valid is $m\sim 10^{-3}\;{\rm{eV}}/{\rm{c}}^2$ (see \S5.3).
This is larger than the largest minimum mass RDS 2011 requires for vortex
formation, $m\simeq 10^{-21}\;\rm{eV}/\rm{c^2}$.\footnote{See their (2011), Table 2; we
  use dwarf-galaxy values as they provide roughly the same radius and mass as the
  degenerate core where we are concerned to determine if vortices form.} Using eq. 
(\ref{eq:rc}) to relate $U_0$ to $m$ and $\rc$, the minimum value of $U_0$ (which
occurs at the minimum mass computed above) is $U_0\simeq 10^{-22}\; \rm{eV}\;\rm{cm^3}$.
This is stronger than the largest minimum required repulsion of
$2\times10^{-58}\;\rm{eV}\;\rm{cm^3}$. Thus, vortices will form as long as the ancillary
assumptions (e.g., non-zero spin parameter $\lambda$) made by RDS 2011
are satisfied.

However, it should be borne in mind that, as we have shown above, were vortices to form, if they depleted the core density enough to evade the conflicting bounds we place on the DM particle's mass, the observationally-required core would disappear.  Thus, to reiterate, they do not affect our conclusion that repulsive bosonic DM in thermal equilibrium is observationally ruled out.  Nonetheless, one might worry that if a vortex forms precisely at the center of the halo, then the scaling we have used throughout, $\hat{\nu}(0)\theta=C\equiv 1$ (see \S3) would be invalid, as the density at the center of the vortex (and hence halo) would be zero.  In response to this, it must be observed that the vortex will be quite small.  Its size is given by the healing length, $\xi$ (see RDS 2011, Fetter \& Foot 2012 for details):
\begin{equation}
\xi=\frac{\hbar}{\sqrt{2U_0\rho}}\approx\frac{\hbar}{mv_{\rm{c}}}\simeq.6\;\rm{m},
\end{equation}
where we have used the definition of $v_{\rm{crit}}$ and then that $v_{\rm{crit}}\simeq v_{\rm{c}}$ to obtain the second and third equalities. Hence any vortex would be minuscule.  In any case, were a vortex present at the center of the halo, one might simply replace our $\hat{\nu}(0)$ with $\left<\hat{\nu}(r<r_s)\right>$, with $r_s$ some smoothing length defining a sphere over which the average density (denoted by the angle brackets) is taken.

\subsection{Drag on a rotating galactic bar}
Finally, we consider the drag on a rotating galactic bar in our model. Since we have
argued that bosonic repulsive dark matter in thermal equilibrium should be ruled out, this
discussion is of secondary importance.  We refer the reader to Appendix B for our treatment.



\section{Acknowledgements}
JG gratefully acknowledges the Chandrasekhar Centenary Conference for the opportunity to present an earlier version of this work; ZS expresses thanks to Dr. Gregory Novak for useful conversations during the course of this research.  ZS also thanks Tanja Rindler-Daller for helpful correspondence on vortices,  David Spergel for discussion on the current state of observational DM research, and Neta Bahcall for intuition on the broad techniques used to determine DM masses.  This material is based upon work supported by the National Science Foundation Graduate Research Fellowship under Grant No. DGE- 1144152. Finally, we are very grateful to the anonymous referee for many careful comments that significantly improved the substance of this work.

\section{References}
\hspace{0.2in} 

Ahn K and Shapiro P, 2005, MNRAS 363, 1092-1124, arXiv:astro-ph/0412169

Amaro-Seoane P, Barranco J, Bernal A, Rezzolla L, 2010, JCAP 1011: 002, arXiv:1009.0019v2 

Annett JF, 2004, Superconductivity, Superfluids, and Condensates, Oxford: University Press

Arbey A, Lesgourgues J, and Salati P, 2003, Phys. Rev. D 68,
023511, arXiv:astro-ph/0301533 

Baldeschi MR, Gelmini GB, and Ruffini R, 1983, Phys. Lett. B 122, 221

Bekenstein JD, 2006, Contemporary Physics 47, 387, arXiv:astro-ph/0701848

Bernal A, Matos T, and Nunez D, 2008, Revista Mexicana de astronomia y astrofisica, 44, 1, arXiv:astro-ph/0303455 

Bernal A, Guzman FS, 2006, Phys. Rev. D 74: 103002, arXiv:astro-ph/0610682 

Bertone G, Hooper D, and Silk J, 2005, Phys. Rept. 405: 279-390, arXiv:hep-ph/0404175v2

Binney  J and Tremaine S, 2008, Galactic Dynamics: Second Edition Princeton: Univ. Pr.

Bohmer CG and Harko T, 2007, JCAP 0706:025, arXiv:0705.4158v4

Brada\u{c} M et al., 2008, ApJ 687 959, http://iopscience.iop.org/0004-637X/687/2/959/

Briscese F, 2011, arXiv:1001.0028v3 

Chavanis PH, 2011, 	Phys. Rev. D 84, 043531, arXiv:1103.2050v2 

Chavanis PH and Delfini L, 2011, Phys. Rev. D 84, 043532, arXiv:1103.2054

Cole DR, Dehnen W, and Wilkinson MI, 2011, MNRAS 16, 2, 1118-1134, arXiv:1105.4050

Colin P, Klypin A, Valenzuela O, and Gottlober S, 2004, ApJ 612, 50, arXiv:astro-ph/0308348 

Colpi M, Shapiro SL, and Wasserman I, 1986, Phys. Rev. Lett. 57, 2485 

Debattista VP and Sellwood JA, 1998, ApJ 493: L5-L8, arXiv:astro-ph/9710039 

Debattista VP and Sellwood JA, 2000, ApJ 543, 704-721, arXiv:astro-ph/0006275 

de Blok WJG, McGaugh S, Bosma A, and Rubin VC, 2001, ApJ 552, L23-26, arXiv:astro-ph/0103102

Elson EC, de Blok WJG, and Kraan-Korteweg RC, 2010, MNRAS 404, 4, 2061-2076, arXiv:1002.0403v1 

Fetter AL and Foot CJ, 2012, arXiv:1203.3183v1

Diemand J, Moore B, and Stadel J, 2005, Nature. 433, 389, arXiv:astro-ph/0501589 

Gerssen J, Kuijken K, and Merrifield MR, 1999, MNRAS 306, 926

Goldreich P and Nicholson PD, 1989, ApJ, 342, 1079

Guaenault AM, 2003, Basic Superfluids, New York: Taylor and Francis

Harko T, 2011, arXiv:1105.5189v1 (II) 

Harko T, 2011, arXiv:1105.2996v1 (II) 

Hennawi J and Ostriker J, 2002, ApJ, 572 41, arXiv:astro-ph/0108203

Hirota A et al., 2009, PASJ, 61, 441

Hu W, Barkana R, and Gruzinov A, 2000, Phys. Rev. Lett. 85,
1158, arXiv:astro-ph/0003365 

Jardel JR and Sellwood JA, 2009, ApJ  691, 1300, arXiv:0808.3449  

Kapusta JI and Gale C, 2006, Finite-Temperature Field Theory, Cambridge: University Press

Kaup DJ, 1968, Phys. Rev. 172, 1331 

Koda J, Shapiro P, 2011, arXiv:1101.3097v2

Komatsu E et al., 2011, ApJS 192: 18, arXiv:1001.4538

Kouvaris C and Tinyakov P, 2011, PRL 107, 091301, arXiv:1104.0382

Lee J and Koh I, 1996, Phys. Rev. D 53, 2236, arXiv:hep-ph/9507385 

Lee J-W, 2009, Journal of the Korean Physical Society, 54, 6, 2622,  arXiv:0801.1442v4 

Lee J-W and Lim S, 2010, JCAP 1001: 007, arXiv:0812.1342v3 

Lin T, Yu H-B,  and Zurek K, 2011, arXiv:1111.0293v1 

Loeb A and Weiner N, 2011, Phys. Rev. Lett. 106: 171302, arXiv:1011.6374v1 

Mandl F, 1988, Statistical Physics, Chichester: John Wiley \& Sons

Merrifield MR and Kuijken K, 1995, MNRAS, 274, 933

Mikheeva E, Doroshkevich A, and Lukash V, 2007, Nuovo Cim. 122B, 1393, arXiv:0712.1688 

Miralda-Escud\'e J, 2002, ApJ 564: 60-64, arXiv:astro-ph/0002050 

Mitra S, 2005, Phys. Rev. D 71, 121302, arXiv:astro-ph/0409121v5

Mo HJ and Mao S, 2000, MNRAS 318, 163, arXiv:astro-ph/0002451v2

Peebles PJE, 1993, Principles of Physical Cosmology, Princeton: University Press

Pethick C and Smith H, 2002, Bose-Einstein Condensation in Dilute Gases, Cambridge: University Press

Pitaevskii L and Lifshitz EM, 1980, Statistical Physics. Part 2. Oxford: Pergamon

Proukakis N and Jackson B, 2008, J. Phys. B: At. Mol. Opt. Phys. 41 203002

Randall SW, Markevitch M, Clowe D, Gonzalez AH, and Bradac M, 2008, ApJ 679,
1173, http://iopscience.iop.org/0004-637X/679/2/1173/fulltext/71772.text.html

Rindler-Daller T and Shapiro P, 2010, New Horizons in Astronomy (Bash Symposium 2009), Proceedings of the Astronomical Society of the Pacific, eds. L Stanford, L Hao, Y Mao, J Green, arXiv:0912.2897v2

Rindler-Daller T and Shapiro P, 2011, arXiv:1106.1256v1

Riotto A and Tkachev I, 2000, Physics Letters B, 484, 177

Ruffini R and Bonazzola S, 1969, Phys. Rev. 187, 1767 

Sahni V, 2004, Proceedings of the Second Aegean Summer School on the Early Universe, Syros, Greece, arXiv:astro-ph/0403324v3

Schunck FE, 1998, arXiv:astro-ph/9802258

Sin SJ, 1994, Phys. Rev. D 50, 3650 

Spergel D and Steinhardt PJ, 2000, Observational evidence for self-interacting cold dark matter,
arXiv:astro-ph/9909386v2

Springel V and Farrar G, 2007, MNRAS 308, 911-925.

Su K-Y, Chen P, 2011, JCAP08, 016, arXiv:1008.3717v2 

Tolman RC, 1938, The Principles of Statistical Mechanics, Oxford: University Press

Tremaine S and Ostriker J, 1999, MNRAS 306, 662, arXiv:astro-ph/9812145

Urena-Lopez L, 2009, JCAP0901: 014, arXiv:0806.3093v2 

Madison KW, Chevy F, Wohlleben W, and Dalibard J, 2000,
Mod. Opt., 47, 2715 

Matos T and Guzman FS, 1999, F. Astron. Nachr. 320, 97

Matos T and Urena-Lopez L, 2002, Phys. Lett. B 538, 246-250, arXiv:astro-ph/0010226 

Wandelt BD, Dav\'{e} R, Farrar GR, McGuire PC, Spergel DN, and Steinhardt PJ, 2000, Proceedings of Dark Matter, arXiv:astro-ph/0006344v2

\section{Appendix A: Equation of state}
\subsection{Partition function to pressure}

Insofar as possible, we follow elementary treatments of a noninteracting ideal boson gas, describing microstates by the
number of quanta $(n_0,n_1,\ldots,n_k,\ldots)\equiv \vec n$ in each single-particle momentum state $\left | k \right>$.
The total number and volume of the gas are $N=\sum n_k$ and $V$, respectively.  The pairwise potential is so short range
as to be represented by a constant, $U_0$, in momentum space.  The energy of microstate $\vec n$, including exchange
terms, is then
\begin{align}
  \label{eq:En}
  E(\vec n)&= \sum_k\frac{k^2}{2m}n_k + \frac{U_0}{2V}\left[N^2 +
\sum_k\sum_{j\ne k} n_jn_k\right]\\
  \label{eq:Enp}
 &= \sum_k\frac{k^2}{2m}n_k + \chi\left[2\sum_k n_k - N^{-1}\sum_k n_k^2\right]\,,
\end{align}
where $\chi\equiv NU_0/2V$ is proportional to the mean potential energy per particle.  The grand-canonical partition
function (GCPF) cannot be found by summing over the $n_k$ independently, even at fixed $\chi$, because of the terms
$\propto -n_k^2/N$.  In the thermodynamic limit, however, only the ground state can have a macroscopic occupation number
(i.e., $\lim_{N\to\infty}\,n_0/N>0$ at fixed $N/V$ and $\beta$), for the usual reasons.  Therefore, $N^{-1}\sum_k n_k^2$ can be
replaced by $N^{-1}n_0^2$.

The canonical (not grand-canonical) partition function is
\begin{equation}
  \label{eq:convolve}
  Z(\beta,V,N)=\sum\limits_{n_0=0}^N Z_0(\beta,V,n_0)Z_{\rm nd}(\beta,V,\,N-n_0),
\end{equation}
in which $Z_0$ and $Z_{\rm nd}$ are the canonical partition functions (CPFs) of the ground single-particle state and of the remaining states,
respectively (``nd''  means ``non-degenerate'').
The summand  eq. (\ref{eq:convolve})  is sharply peaked either at $\bar n_0=0$, or at $n_0=\bar n_0$ such that
\begin{equation}
  \label{eq:peak}
  \left.\frac{\partial\ln Z_{0}}{\partial n_0}\right|_{n_0=\bar n_0}=
  \left.\frac{\partial\ln Z_{\rm nd}}{\partial N_{\rm nd}}\right|_{N_{\rm nd}=N-\bar n_0}\,,
\end{equation}
if this has a solution for $\bar n_0>0$.  Hence in the thermodynamic limit,
\begin{equation}
  \label{eq:Zsum}
 \ln Z(\beta,V,N)= \ln Z_0(\beta,V,\bar n_0) + \ln Z_{\rm nd}(\beta,V,\,N-\bar n_0).
\end{equation}

It remains to find $Z_0$ and $Z_{\rm nd}$.  Neglecting $p_0^2/2m= O(V^{-2/3}\hbar^2/m)$,

\begin{equation}\label{eq:Zground}
Z_0(\beta,V,n_0)=\exp\left[-\beta\chi\left(2n_0-N^{-1}n_0^2\right)\right].
\end{equation}

Note that $Z_0$ depends implicitly on $N$ through $\chi$ as well as $N^{-1}$.  

To obtain $Z_{\rm nd}$, we begin by finding the grand-canonical partition function (GCPF) for the non-degenerate states, which we denote $\mathcal{Z}_{\rm nd}$; it will lead to the canonical partition function $Z_{\rm nd}$ via eq. (\ref{eq:Znd}).  Because the $n_k^2$ terms have
been dropped for $k>0$, it can be computed as for an ideal gas, with modal
energies $\epsilon_k = p_k^2/2m + 2\chi$:
\begin{eqnarray}
\ln\mathcal{Z}_{\rm nd}(\beta,V,\mu_{\rm nd})  &=&  \sum_{k=1}^\infty \ln\left(\sum_{n_k=0}^\infty e^{-n_k\beta(\epsilon_k-\mu_{\rm nd})}\right)\nonumber\\
&\approx& V\nu_{\rm crit}(\beta)\zeta(\tfrac{3}{2})^{-1}\textrm{Li}_{5/2}\left[e^{\beta(\mu_{\rm nd}-2\chi)}\right].
  \label{eq:gcpf}
\end{eqnarray}

The polylogarithm $\textrm{Li}_{5/2}$, which is equivalent to a Bose-Einstein integral,
results from approximating the sum over $k$ by an integral over $p_k$.  The critical density $\nu_{\rm crit}(\beta)$
is defined in eq. (\ref{eq:nucrit}). 

The quantity $\mu_{\rm nd}$ is not the chemical potential of the full system, which would be conjugate to $N$,
but instead describes the division of particles between the ground state and the rest:
\begin{subequations}
\begin{align}
  \label{eq:mund}
 \bar N_{\rm nd}&= \left(\frac{\partial \mathcal{\ln Z}_{\rm nd}}{\partial\mu_{\rm nd}}\right)_{\beta,V,N} = N-\bar n_0,\\[1ex]
  \label{eq:muofN}
  \mu_{\rm nd} &= -\beta^{-1}\left(\frac{\partial\ln Z_{\rm nd}}{\partial \bar N_{\rm nd}}\right)_{\beta,V,N}\,,
\end{align}
\end{subequations}
both of which are implied by the thermodynamic relation
\begin{equation}
  \label{eq:Znd}
\ln Z_{\rm nd}(\beta,V,\bar N_{\rm nd})=\ln \mathcal{Z}_{\rm nd}(\beta,V,\mu_{\rm nd}) - \beta \mu_{\rm nd}\bar N_{\rm nd}\,.
\end{equation}

Recall that $\mathcal {Z}_{\rm nd}$ is the GCPF, not the CPF, but we require the CPF $Z_{\rm nd}$. Here, the GCPF is only a device for obtaining the CPF.

Combining eqs. (\ref{eq:peak}), (\ref{eq:Zground}), and (\ref{eq:muofN}) yields
\begin{equation}
  \label{eq:muofN2}
  \mu_{\rm nd} = 2\left(1-\frac{\bar n_0}{N}\right)\chi\,,
\end{equation}
but only when $\bar n_0$ is macroscopic, since otherwise eq. (\ref{eq:peak}) does not hold.

Combining eqs. (\ref{eq:Zsum}), (\ref{eq:Zground}),
(\ref{eq:gcpf}), and (\ref{eq:Znd}):
\begin{multline}
  \label{eq:Zcantot}
V^{-1} \ln Z(\beta,V,N) = -\beta\chi\left(2\nu_0-\frac{\nu_0^2}{\nu}\right) -\beta\mu_{\rm nd}(\nu-\nu_0) \\
+\nu_{\rm crit}(\beta)\zeta(\tfrac{3}{2})^{-1}\textrm{Li}_{5/2}\left[e^{\beta(\mu_{\rm nd}-2\chi)}\right]\,,
\end{multline}
where the intensive variables $\nu\equiv N/V$ and $\nu_0\equiv\bar n_0/V$ have been introduced.
In this global CPF,
$\mu_{\rm nd}$ and $\nu_0$ are  functions of $(\beta,V,N)$ given by
eqs. (\ref{eq:mund}) and (\ref{eq:muofN2}) if $\bar n_0$ is macroscopic; else
$\nu_0\to 0$ and $\bar N_{\rm nd}\to N$ in eq. (\ref{eq:mund}).
More explicitly,
\begin{align}
  \label{eq:single}
      \nu&= \displaystyle\nu_0 + \nu_{\rm crit}(\beta)\zeta(\tfrac{3}{2})^{-1}\textrm{Li}_{3/2}
\left[e^{-\beta U_0\nu_0}\right]& \nu>\nu_{\rm crit}\,,\nonumber\\
\nu&=\displaystyle\phantom{\nu_0 +}
\nu_{\rm crit}(\beta)\zeta(\tfrac{3}{2})^{-1}\textrm{Li}_{3/2}\left[e^{\beta[\mu_{\rm nd}-U_0\nu]}\right]&\nu\le\nu_{\rm crit}.
\end{align}

The pressure is now computable as
\begin{align}
  \label{eq:PZ}
  P&=\beta^{-1}\left(\frac{\partial\ln Z}{\partial V}\right)_{\beta,N}\nonumber\\
&=\beta^{-1}\left[\left(\frac{\partial\ln Z_0}{\partial V}\right)_{\beta,\bar n_0}+
\left(\frac{\partial\ln Z_{\rm nd}}{\partial V}\right)_{\beta,N-\bar n_0}\right]\,.
\end{align}
In the second line, $\bar n_0$ can be treated as constant because
the terms involving $\partial\bar n_0/\partial V$ cancel from  eq. (\ref{eq:PZ}) due to eq. (\ref{eq:peak}).
Similarly, terms involving $\partial\mu_{\rm nd}/\partial V$ cancel from the derivative of
$Z_{\rm nd}$ in the form of eq. (\ref{eq:Znd}) due to eq. (\ref{eq:mund}).  But terms involving
$\partial\chi/\partial V$ do matter.
After some manipulation,
\begin{equation}
  P= U_0\left(\nu^2-\tfrac{1}{2}\nu_0^2\right)+
\beta^{-1}\nu_{\rm crit}(\beta)\zeta(\tfrac{3}{2})^{-1}\textrm{Li}_{5/2}\left[e^{\beta[\mu_{\rm nd}-U_0\nu]}\right].
\label{eq:pressure}
\end{equation}

\subsection{Limiting behavior}

In this section, we show that the approximate equation of state defined by
eqs.~\eqref{e:eos}-\eqref{e:zofnu0} has the expected behavior at both extremes of the
degeneracy parameter $\nu/\nu_{\rm crit}(T)$.  The Hartree-Fock approximation is that the
boson gas can be described in terms of occupation numbers of single-particle states, and
the discussion below reflects this viewpoint---which, however, is not accurate at
intermediate densities where quasiparticles dominate, as discussed in
\S\ref{subsec:pressure}.

When $T=0$, all of the DM particles will be in the lowest-energy state, corresponding to
a pure Bose-Einstein condensate.  This is evident from the fact that
$\nu_{\rm{crit}}\propto T^{3/2}$, so that $\nu_{\rm{crit}} \rightarrow 0$ as $T\rightarrow 0$.
The equation of state should reduce to the one given in \S2.1,
$P=K\rho^2=U_0\nu_0^2/2$, which follows from the fact that there are $n^2/2$ pairs per
unit volume and that the energy per pair is $U_0\delta(\bs{r}_1-\bs{r}_2)$.
The constant $U_0$ in $P$ corresponds to the ``bare'' interaction only to first order; at
higher orders in perturbation theory it must be renormalized.

In the argument of the polylogarithm in eq. (\ref{eq:pressure}),
$\mu_{\rm{nd}}-U_0\nu$, with $\nu=\nu_0$, simplifies to $-U_0\nu\bar{n}_0/N$, clearly
independent of $T$.  Thus that quantity remains negative as $T\rightarrow 0$.
$\beta\rightarrow\infty$ as $T\rightarrow 0$, so the argument of the polylogarithm in
eq. (\ref{eq:pressure}) becomes zero.  The term multiplying it,
$\beta^{-1}\zeta(\tfrac{3}{2})$, also goes to zero.  So the pressure becomes
$P=U_0\left(\nu^2-\tfrac{1}{2}\nu_0^2\right)$. Since $\nu=\nu_0$ because all of the DM is
in the condensate, this reduces to $P=U_0\nu_0^2/2$ as required.

In the limit that $U_0\rightarrow 0$, the gas is non-interacting and should approach an
ideal gas of indistinguishable bosons. For a given $\nu$, if $T$ is such that $\nu \leq
\nu_{\rm{crit}}$, there will be no condensate, and the chemical potential $\mu_{\rm{nd}}$,
which governs particles' transition from the ground state to excited states, is determined
by inverting eq. (\ref{eq:single}) ($\nu\leq\nu_{\rm{crit}}$).  Taking $U_0=0$ and
rewriting the second term in eq. (\ref{eq:pressure}) using eq. (\ref{eq:single})
($\nu\leq\nu_{\rm{crit}}$), we have
\begin{equation}
\label{eq:pressureU0zeroThighlim}
P=\beta^{-1}\nu\frac{\Li_{\rm{5/2}}\left[e^{\beta\mu_{\rm{nd}}}\right]}{\Li_{\rm{3/2}}\left[e^{\beta\mu_{\rm{nd}}}\right]}.
\end{equation}
This agrees with standard results (e.g. Tolman 1938). For an interacting gas in the dilute limit where the dimensionless combination $\beta
U_0\nu\to 0^+$, the same result \eqref{eq:pressureU0zeroThighlim} still obtains (note that $U_0$ has dimensions of
energy times volume). 

Now we consider $T$ such that $\nu>\nu_{\rm{crit}}$.  In this case a condensate will be
present, so we can use eq. (\ref{eq:muofN}) for $\mu_{\rm{nd}}$. With $U_0=0$,
$\mu_{\rm{nd}}=0$.  As $T\to 0^+$,
the argument of the polylogarithms in eqs.
(\ref{eq:single}) ($\nu>\nu_{\rm{crit}}$) and (\ref{eq:pressure}) approaches unity, and
the functions approach Riemann zeta functions.  We now replace the second term in eq.
(\ref{eq:pressure}) using eq. (\ref{eq:single}) ($\nu > \nu_{\rm{crit}}$).  These
manipulations yield
\begin{equation}
P=\beta^{-1}\left(\nu-\nu_0\right)\frac{\zeta\left(\tfrac{5}{2}\right)}{\zeta\left(\tfrac{3}{2}\right)}.
\label{eq:pressureUeq0andcond}
\end{equation}

Note that this result is what we would obtain if in our eq.
(\ref{eq:pressureU0zeroThighlim}) (or Tolman's 93.20) we took $\mu_{\rm{nd}}=0$ and
replaced $\nu$ with $\nu-\nu_0$, which is justified because normally when a condensate
appears the chemical potential in the non-degenerate component is set to zero and only the
non-degenerate component produces pressure.  This is because particles in the condensate
have zero momentum down to quantum mechanical uncertainty ($\Delta p_{0,\:\rm{min}} \simeq
V^{-1/3}\hbar$) and so should not contribute any pressure.

\section{Appendix B: Dynamical Friction}

Goodman (2000) argued that the superfluidity of an RDM condensate implies that there
should be no dynamical friction on a moving potential, such as a galactic bar, insofar as
the core of the halo is dominated by the condensate and $v_{\rm bar}<v_{\rm
  crit}=\sqrt{2\nu U_0/m}$.  Our current view is somewhat different.  In the limit that
the scattering mean-free-path is large compared to the dimensions of the bar or the core,
the original argument is perhaps correct.  In the opposite limit, while it remains true
that the dynamical friction vanishes below a critical speed $\sim v_{\rm crit}$, the
reason is not because the RDM is a superfluid (although it is), but rather because it is
sufficiently ideal fluid in the classical sense, i.e. the viscosity due to collisions
between particles and due to the thermal velocity of the particles is negligible.

The important difference between a galactic bar interacting with DM gravitationally and a
spoon dragged through a laboratory fluid is that the gravitational potential is smooth,
whereas the spoon has a surface at which a no-slip boundary condition applies to the
normal (non-superfluid) component.  Thus, even in laminar flow, the spoon exerts a viscous
force on the normal component that vanishes more slowly than linearly with the viscosity,
probably as the Reynolds number $Re^{-1/2}$, because there is a laminar boundary layer.
At moderately high Reynolds number, the boundary layer on the rear side of the spoon
becomes unstable and contributes a turbulent drag that is asymptotically independent of
$Re$.  For a smooth large-scale potential like that of a rotating bar, there is no such
surface and no such boundary layer, and therefore perhaps no turbulence.  This absence of
drag would seem to hold even if the RDM had a substantial normal-fluid component.

However, the bar or other moving potential may experience a wave drag if it
couples to a wave whose phase velocity, whether linear or angular, matches that of the
potential itself.  In a pure condensate, below $v_{\rm crit}$, the only significant waves
are phonons at speed $\cs=\sqrt{2K\rho}$, so that a
linearly moving potential experiences no wave drag if its velocity is subsonic, though a
rotating bar could excite phonons at large radii where $|\bs{r\times\Omega}_{\rm
  bar}|>\cs$.  One can see this explicitly from the quantum-mechanical equation of
motion for the condensate wave function (Gross-Pitaevskii or non-linear Schr\"odinger
equation):
\begin{equation}
  i\hbar\frac{\partial\Psi}{\partial
    t}=-\frac{\hbar^2}{2m}\nabla^2\Psi+U_0|\Psi|^2\Psi+(\Phi_{\rm self}+\Phi_{\rm ext})\Psi\,,
      \label{eq:NLSE}
\end{equation}
in which $\Phi_{\rm ext}$ \& $\Phi_{\rm self}$ are the gravitational potentials of the bar
or other external perturber and of the condensate itself; $\Phi_{\rm self}$ satisfies
Poisson's equation with the mass density $\rho_{\rm c}=m|\Psi|^2$ of the condensate as
source.  

Adopting the usual Jeans swindle, consider perturbations to a background state of
uniform density and vanishing $\Phi$.  The unperturbed wavefunction $\Psi_0$ has a
constant modulus $|\Psi_0|=\sqrt{\nu_0}$ but a time-dependent phase, because the
term involving $U_0$ in eq. (\ref{eq:NLSE}) does not vanish in the background state.  One
sets $\Psi\to\Psi_0(t)[1+\varepsilon(\bs{r},t)]$ and expands eq. (\ref{eq:NLSE}) to first
order in the real and imaginary parts of $\varepsilon$, treating $\Phi_{\rm ext}$ and
$\Phi_{\rm self}$ as well as $\varepsilon$ as first-order quantities.  In the unforced case
$\Phi_{\rm ext}=0$, the dispersion relation for Fourier modes
$\varepsilon(\vec{r},t)\propto\exp(i\vec{k}\cdot \vec{r}-i\omega t)$ becomes
\begin{equation}
  \omega_{k}^2=\cs^2k^2-4\pi Gm\nu +\left(\frac{\hbar k^2}{2m}\right)^2\,,
  \label{eq:dispersion}
\end{equation}

in which $\cs\equiv\sqrt{\nu U_0/m}$ plays the role of sound speed.  Chavanis (2011) also
obtains this result.  As usual with the GP Equation, the last term on the right represents
single-particle excitations; it is small for long-wavelength modes $k\ll
2 m\cs /\hbar$.  If one neglects the term in $k^4$, then equation
(\ref{eq:dispersion}) matches the results of a Jeans analysis for a classical ideal fluid,
$\omega^2=c_s^2 k^2-4\pi G\rho$ (e.g. Binney \& Tremaine 2008).

This conclusion can also be reached if one explicitly computes the drag force on a
perturbing body.  Consider a rigid perturbing potential that moves at constant velocity
$v$ through the condensate, $\Phi_{\rm ext}(\vec{r}-\vec{v}t)$.  If we write
$\tilde\Phi_{\rm ext}(\vec{k})$ for the spatial Fourier transform of this potential at any
time, then after transients have decayed, the component of the wave drag along $\vec{v}$
is
\begin{equation}
  F_{\rm drag}=-
  \frac{\rho_0}{V}\int\frac{d\vec{k}}{(2\pi)^3}\,\delta(\omega_{k}-\vec{k}\cdot
    \vec{v})|\vec{k}\tilde\Phi_{\rm ext}(\vec{k})|^2,
      \label{e:drag}
\end{equation}
with $\rho_0\equiv m\nu$.
Since the Fourier components $\tilde\Phi_{\rm ext}(\vec{k})$ are negligible at $k\gtrsim\sqrt{2m\cs}/\hbar$,
this is effectively the same drag as for an ideal fluid with equation of state
$P=K\rho^2$.  The quantum-mechanical nature of the condensate plays no direct role.  
To the extent that the self-gravity of the RDM is slight on the scale of the perturber,
in other words  $\tilde\Phi_{\rm ext}(\vec{k})$ is unimportant for $k^2\lesssim4\pi
G\rho_0/\cs^2$,  the drag vanishes for subsonic motion relative to the condensate.

The formal result \eqref{e:drag} holds more generally for ideal fluids in which the
dispersion relation may differ from eq. (\ref{eq:dispersion}).  Thus it should hold
even when the RDM has a normal (non-degenerate) component, as it must at finite
temperature.  In a realistic case where the background state is not uniform, however, the
RDM gas would be stratified, i.e. it would have an entropy gradient parallel to the
background gravitational field, so that waves restored by buoyancy (internal waves/
g~modes) might be excited at subsonic velocities.  As we show in \S3, an isothermal
RDM halo at nonzero temperature consists of a core that is almost pure condensate, and has
nearly the $n=1$ Emden profile, surrounded by an extended non-degenerate ``atmosphere,''
with a sharp cliff in the density profile at the edge of the core (Fig.~\ref{f:densityprof}).  Although we do not
calculate it here, the coupling of a sub-sonicly rotating bar potential to the g~modes
would probably occur mainly just outside the core, with a drag
proportional to the density in the non-degenerate component there.  Since that density is
much less than the central density, the
drag on the bar for a given mass-to-light ratio would be much less than what Debattista \&
Sellwood (2000) estimate for collisionless dark matter.  The pattern speeds of their
simulated bars slow significantly from their maximum possible values---the values at which
corotation with the local galactic circular velocity occurs near the end of the bar---in a
few rotation periods.  

 Typical pattern speeds of galactic bars are measured to be $\Omega_{\rm
  b}\lesssim 60\,{\rm km\,s^{-1}\,kpc^{-1}}$, corresponding to rotation periods
$2\pi\Omega_{\rm b}^{-1}\gtrsim 10^{8}\,{\rm yr}$ (Merrifield \& Kuijken 1995; Gerssen et
al. 1999; Hirota et al. 2009 and references therein).  An $n=1$ Emden polytrope with
central sound speed $\cs(0)$, radius $\rc$, and mass $M_{\rm c}$ has a circular velocity
$\vc\equiv\sqrt{GM_{\rm c}/\rc} =\cs(0)$.  The slowest (fundamental) mode of such a
polytrope with the required quadrupolar symmetry ($\ell=m=2$) has pattern speed
$\omega_{\rm c}/2 =0.616 \vc/\rc$, which is about the same as the measured bar speeds if
$\vc=100\,{\rm km\,s^{-1}}$ and $\rc=1\,{\rm kpc}$.  If $\Omega_{\rm b}<\omega_c/2$, then
the bar should not excite this mode, and there should be no torque between the bar
and the core.  There would probably still be a drag on the surrounding non-degenerate
component, but without attempting to calculate this explicitly, we expect by analogy with
tidal excitation of g~modes in stars (e.g. Goldreich \& Nicholson 1989) that the torque on
that component would be suppressed by the ratio of the maximum nondegenerate density to
the central density of the core, a factor $\sim\theta\ll 1$.

\end{document}